\documentclass[tighten, twocolumn]{aastex631} 

\begin{document}

\title{Deneb is a Large Amplitude Polarimetric Variable}

\author[0000-0003-0340-7773]{Daniel V. Cotton}
\affiliation{Monterey Institute for Research in Astronomy, 200 Eighth Street, Marina, CA 93933, USA.}

\affiliation{Western Sydney University, Locked Bag 1797, Penrith-South DC, NSW 2751, Australia.}

\author[0000-0002-5726-7000]{Jeremy Bailey}
\affiliation{School of Physics, University of New South Wales, Sydney, NSW 2052, Australia.}

\affiliation{Western Sydney University, Locked Bag 1797, Penrith-South DC, NSW 2751, Australia.}

\author[0000-0002-6703-5406]{Jean Perkins}
\affiliation{Monterey Institute for Research in Astronomy, 200 Eighth Street, Marina, CA 93933, USA.}

\author[0000-0002-1988-143X]{Derek L. Buzasi}
\affiliation{Department of Chemistry \& Physics, Florida Gulf Coast University, 10501 FGCU Boulevard S., Fort Myers, FL 33965, USA.}

\author[0009-0005-6387-6936]{Ievgeniia Boiko}
\affiliation{Monterey Institute for Research in Astronomy, 200 Eighth Street, Marina, CA 93933, USA.}

\affiliation{California State University, Long Beach, 1250 Bellflower Blvd, Long Beach, CA 90840, USA.}

\begin{abstract}
We write to report the discovery that Deneb is a large amplitude polarization variable. Over a $\sim$400~d time span from August 2022 Deneb's polarization was typically around 3900 parts-per-million (ppm) in the SDSS $g^\prime$-band.  Yet, it varied by several hundred ppm in an irregular way on a timescale of weeks. The largest polarization change, amounting to 2500~ppm, occurred shortly after the last pulsation ``resumption'' event identified by \citet{Abt23} in \textit{TESS} photometry. The relationship between the observed polarization -- particularly corresponding to the resumption event -- and its brightness and H$_{\rm \alpha}$ spectra suggests a mechanism involving density changes in its wind and/or extended atmosphere. Smaller effects due to pulsations are not ruled out and further study is recommended.
\end{abstract}


\section{Introduction} \label{sec:intro}

Deneb ($\alpha$ Cygni, HD\,197345) is a type A2\,Ia supergiant, and the 19th brightest star in the night sky ($m_V=1.25$). It has long been known to be variable photometrically ($\Delta m\approx 0.08$, \citealp{Fath35}) and in radial velocity (RV) ($\Delta v= 15.5$ km/s, \citealp{Paddock35}) on similar timescales. \citet{Abt57} found Deneb's behaviour to be typical of intermediate type supergiants, and concluded radial pulsations were the most likely cause. In re-examining \citeauthor{Paddock35}'s data, \citet{Lucy76} identified many periods, and attributed its stable semi-regular variability to the simultaneous excitation of many discrete non-radial pulsation modes. Though sparse, other studies have found inconsistent periods in Deneb's RV and photometry. For example \citet{Richardson11} identified transient periods of $\approx$40~d in H$_\alpha$ structure but 13.4 and 17.8~d in photometry and Si\,\textsc{ii} RV. 

Last year \citet{Rzaev23}, by applying line profile analysis, concluded periods of 12-14~d and $\sim$22~d to be due to radial and non-radial pulsations respectively. Even more recently \citet{Abt23} identified a pattern for resumptions of the 11-12~d pulsation cycle in flux and RV, where they are triggered irregularly at multiples of 72.4\,$\pm$\,0.3~d -- according to \citet{Rzaev23} the events last $\approx$34$\,\pm$\,1~d. Mode identification would enable asteroseismology, but this has proved difficult for other massive non-radial pulsators, even combining spectroscopy and photometry. 

Many other early-type supergiants were added to the $\alpha$~Cygni variable class, of which Deneb is the prototype, after the precise space-based photometry of the \textit{Hipparcos} mission revealed their variability \citep{Adelman97, Waelkens98}. The presumed mechanism, non-radial pulsations, is written into the class definition\footnote{See `ACYG' in the General Catalogue of Variable Stars (\citealp{Samus17}, http://www.sai.msu.su/gcvs/gcvs/iii/vartype.txt).}. However, more recent high-cadence space photometry reveals supergiant variability to be stochastic in nature with a red-noise type frequency distribution. Proposed mechanisms for which include internal gravity waves \citep{Bowman19}, sub-surface convection regions \citep{Cantiello21}, and instabilities in their strong radiatively-driven stellar winds \citep{Krtivcka21}.

Linear polarimetry provides spatial information, either in the form of normalized Stokes vectors $q$ and $u$, or in total linear polarization, $p$, and position angle, $\theta$. It can be used in asteroseismology to break degeneracies and enable mode determination \citep{Cotton22a}. However, large polarization variability has been observed in other $\alpha$ Cygni stars, and explained as arising from scattering in a clumpy wind (e.g. \citealp{Hayes84, Hayes86}, see also \citealp{Clarke10}, Ch.\,13.6). The two mechanisms can be distinguished by the scale and character of the variability. 

The most recent linear polarization observations of Deneb are searches for line polarization producing non-detections (refs. within \citealp{Clarke84}). Broadband observations, typically better constrained, were made from 1949 to $\sim$1964 (Tbl.\,\ref{tab:historic}); most extensively by Alfred Behr, who observed first unfiltered \citep{Behr59a} then in three passbands \citep{Behr59b}. The latter display a significant $\theta$ rotation with wavelength, ascribed to multiple line-of-sight dust clouds. In every case Deneb's polarization is taken as interstellar, with no variability ever claimed.

We first observed Deneb as part of an, as yet, unpublished polarimetric survey of bright northern stars; follow-up observations showed obvious variability. This became part of the impetus for a large study of $\alpha$ Cygni stars as a class by teams in both hemispheres, a report of which will be given in a future publication. In light of \citeauthor{Abt23}'s call for more spectroscopy and photometry of Deneb, we present our current polarimetric data on the star, along with some pertinent calculations, to demonstrate the value polarimetry has for this work.

\begin{table}[]
\centering
\tabcolsep 2.5 pt
\begin{tabular}{lcrrrc}
\hline
Date            &   \multicolumn{1}{c}{$\lambda_{\rm eff}$}  &   $n$   &   \multicolumn{1}{c}{$p$} &   \multicolumn{1}{c}{$\theta$}    &   Ref.  \\
                &   \multicolumn{1}{c}{(nm)}                &         &     \multicolumn{1}{c}{(ppm)}               &   \multicolumn{1}{c}{($^\circ$)}  &   \\
\hline
1949 11/29--12/15 &440  &   4   &   5390\,$\pm$\,1250 &   19\phantom{.0}\,$\pm$\,7\phantom{.0}   &   1,2 \\
1956--1958      &   462 &   6   &   3969\,$\pm$\,\phantom{0}100   & 29\phantom{.0}\,$\pm$\,1\phantom{.0}   &   3 \\
1958 N. Summer  &   372 &   6   &   4802\,$\pm$\,\phantom{0}148   & 31.8\,$\pm$\,1.3 & 4 \\
1958 N. Summer  &   430 &   7   &   4410\,$\pm$\,\phantom{0}100   & 37.7\,$\pm$\,1.3 & 4\\
1958 N. Summer  &   516 &   5   &   4459\,$\pm$\,\phantom{0}123   & 40.2\,$\pm$\,1.3 & 4\\
$c$1964         &   580 &   1   &   3969\,$\pm$\,\phantom{0}196   & 40.2\,$\pm$\,1.4 & 5 \\
\hline
\end{tabular}
\caption{Historic observations of Deneb's polarization. Values of total linear polarization, $p$, given in parts-per-million, are converted from polarization magnitudes. The third column indicates the number of individual observations. Regrettably, only \citet{Hall50} report the individual dates and measurements, but large nominal errors mean these are not illuminating. 
No correction has been made for co-ordinate precession of $\theta$, which is $\approx0.01^\circ$/yr. 
Refs: 1: \citet{Hall50}, 2: \citet{Hall58}, 3: \citet{Behr59a}, 4: \citet{Behr59b}, 5: \citet{Serkowski69}. \label{tab:historic}}
\end{table}

\section{Observations} \label{sec:obs}
From 2022 August to 2023 October, high precision polarimetric observations of Deneb were made with the HIgh Precision Polarimetric Instrument 2 (\mbox{HIPPI-2}, \citealp{Bailey20}) on MIRA's 36-inch telescope at its Oliver Observing Station (OOS) \citep{Cotton22b}, and with the Polarimeter using Imaging CMOS Sensor And Rotating Retarder (PICSARR, \citealp{Bailey23a}), first on the Celestron C14 at MIRA's Weaver Student Observatory (WSO, \citealp{Babcock08}) and later on a 14-inch CDK co-mounted to the 36-inch. Most observations were 12~min in the $g^\prime$ band (errors: 7, 17, 17 ppm respectively, i.e. $\sim$10-30$\times$ more precise than the best previous individual measurements); some shorter exposures were made in other SDSS bands with PICSARR. Data were reduced by the usual calibration and reduction procedures of each instrument, involving low/high polarization standard observations and a full bandpass model (without reddening). Multi-band observations of polarized standards are used to correct $\theta$. 

During this period, 8 spectra (SNR $\sim$150-200) of Deneb over 6~nights were taken with a BACHES Mini-echelle spectrograph (R$\sim$35,000) on the MIRA 36-inch. These were reduced with standard IRAF tools and procedures.

The \textit{Transiting Exoplanet Survey Satellite} (\textit{TESS}, \citealp{Ricker14}) observed Deneb in Sectors 41, 55 and 56 at 120~s cadence. We used the Asteroseismology-Optimized Pipeline (AOP) software to align and process the photometric data \citep{Buzasi2016, Nielsen20, Metcalfe2023}. This essentially constructs an optimized photometric aperture one pixel at a time, minimizing the high-frequency noise in the resulting light curve, which is then iteratively detrended against centroid position and background -- the result is a light curve that is consistently significantly better than the SPOC product for bright and/or saturated targets.

\section{Analysis and Discussion} \label{sec:analysis}

\subsection{Mean Polarization and Variability}

\begin{figure*}
    \includegraphics[width=\textwidth, trim={0 0.275cm 0 0.2cm}, clip]{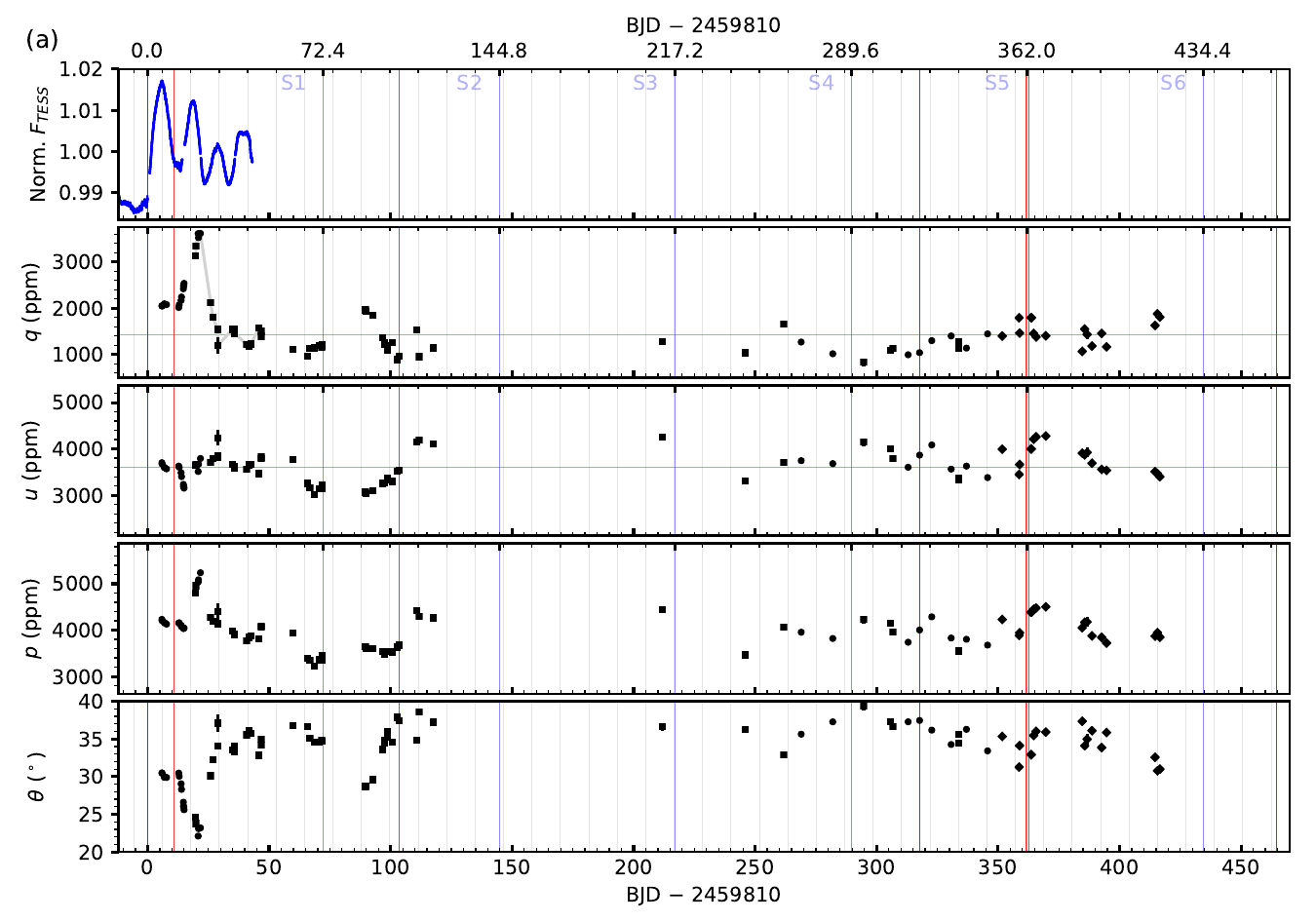}
    \includegraphics[angle=0, width=0.55\textwidth, trim={-2.0cm 0cm -0.5cm 2cm}, clip]{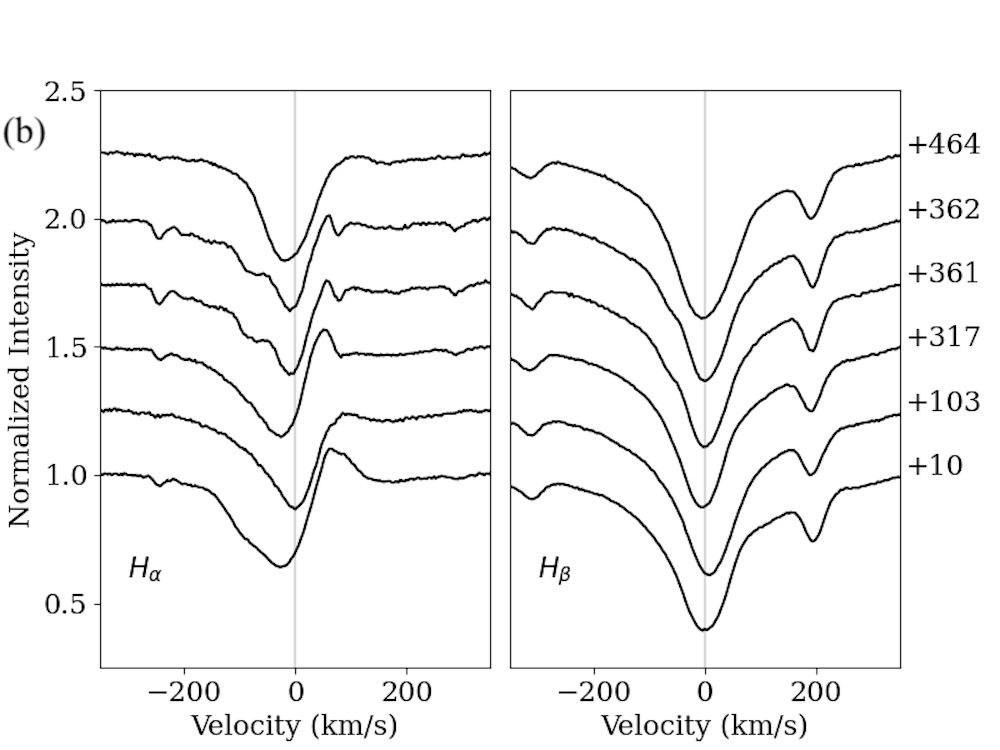}
    \includegraphics[width=0.40\textwidth, trim={-1cm, -5cm, -1cm, 0}, clip]{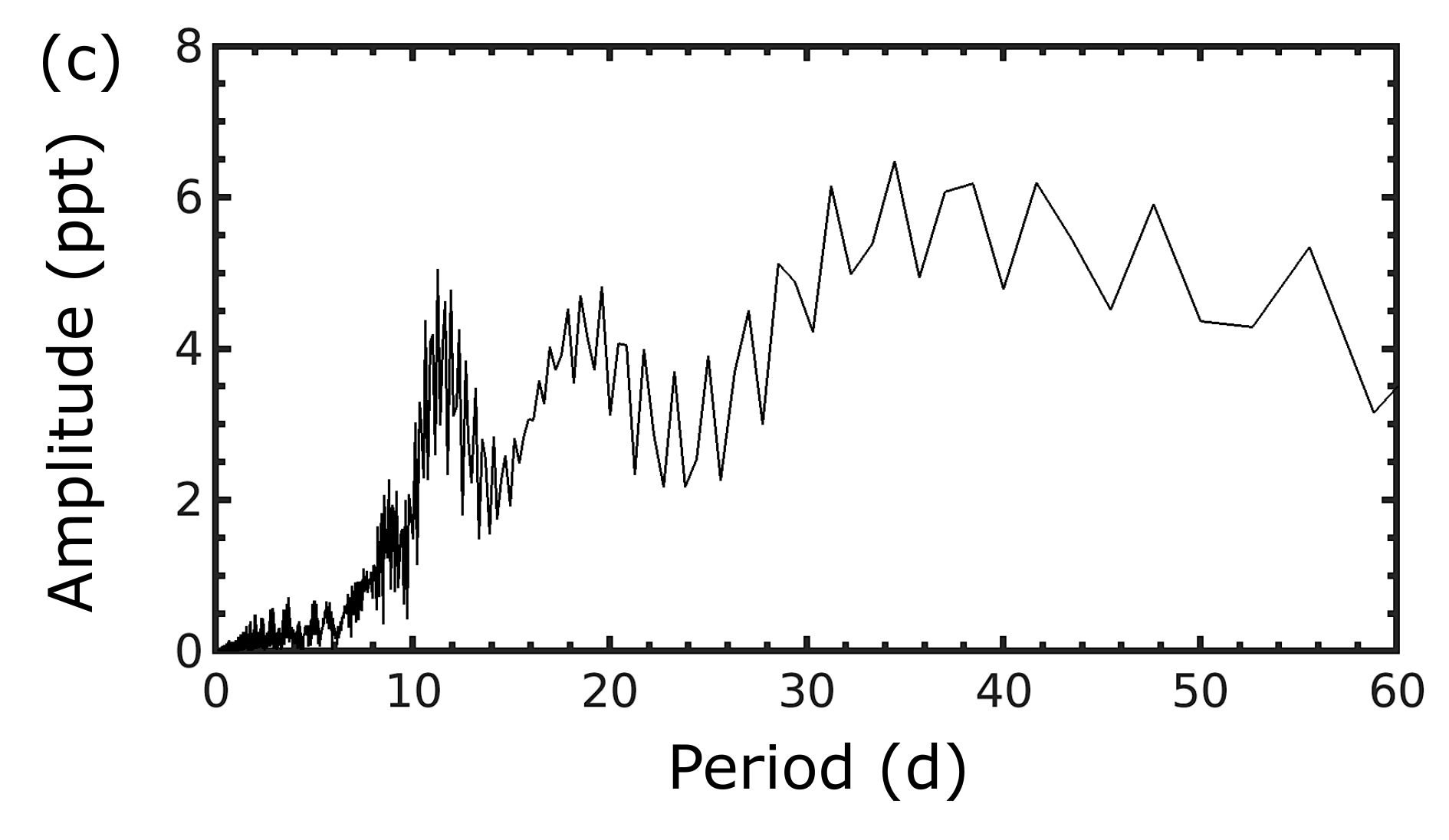}
    \caption{Observational data. (a) Time series data from Sector 55 and 56 \textit{TESS} photometry [top panel] and $g^\prime$-band polarimetry (HIPPI-2/36-inch/OOS -- circles, PICSARR/C14/WSO -- squares, PICSARR/CDK-14/OOS -- diamonds) [bottom four panels]. The time series are divided into a number of segments, which we denote S[n]; these are all 72.4~d in length corresponding to the semi-regular cycle for the ``resumption'' events. The beginning of the first segment is 2459810 BJD, which is the last resumption event identified by \citet{Abt23}. Deneb's 11.7~d pulsation cycle can be seen in the \textit{TESS} data; the grey grid lines are placed every 11.7 days offset to $+$6.1~d to correspond with the first photometric maximum. Green lines are the medians in $q$ (1418.5~ppm) and $u$ (3612.5~ppm). A grey line guides the eye in $q$ after the extreme event. The red lines correspond to the BJD dates BACHES spectra (SNR $=$ 150-200) were taken; these are shown in (b) centred on H$_{\rm \alpha}$/H$_{\rm \beta}$ (left/right); the strongest additional absorption features are seen at $+$10, $+$361 and $+$362~d. (c) (Semi-) Amplitude spectrum derived from \textit{TESS} Sectors 41, 55 and 56 in parts-per-thousand (ppt $=1000\times$~ppm); prominent peaks correspond to periods of $\sim$38~d, 11-12~d, and 18-20~d. Note that the frequency resolution is low due to the short length of the time series. 
    \label{fig:gts}}
\end{figure*}

The mean polarization of all 88 $g^\prime$ observations is $p=$ 3947.5~ppm at a position angle $\theta=$ 33.07$^\circ$, or in normalized Stokes parameters: $(q, u)=$ (1596.2, 3610.4)~ppm. In Fig.\,\ref{fig:gts}(a) measurements from the two polarimeters agree well in recording polarization varying on a timescale of weeks. The variability is much larger than the median error of 16~ppm, with $(\sigma_q, \sigma_u)=$ (608.5, 318.9)~ppm, thus $\sigma_p=(\sigma_q^2+\sigma_u^2)^{1/2}=$\,687.0~ppm. Tbl.\,\ref{tab:historic} shows that such variability has been detectable since the second half of the 20th century. Indeed, taking account of wavelength dependence (see Sec.\,\ref{sec:multi}), comparison with Tbl.\,\ref{tab:historic} implies similar past variability. The impression is strengthened by also considering the 18 individual H$_{\rm \beta}$, H$_{\rm \gamma}$ and Ca\,\textsc{ii}\,H line and adjacent continuum observations of \citet{Hayes74,Hayes75,Clarke76,Clarke84}\footnote{In some of this work Deneb is used as a stable reference to check the reliability of the instrument.}; these mostly have errors $\sim$300~ppm, but cover a range $p=$ 3670 to 5400~ppm, and $\theta=$ 32.5 to 42.2$^\circ$. 

Applying the methods of \citet{Brooks94}, both the kurtosis (2.6490) and skewness (5.5924) of $q$ are significant at the 99\% level (but insignificant in $u$ at 0.0454 and 2.4200 respectively). This reflects the irregular nature of the variability recorded, with one deviation from the mean twice the magnitude of any others. There are no obvious periods but changes equating to hundreds of ppm are seen to occur on a timescale of weeks. Together with the sparsity of observations, this explains how Deneb's polarimetric variability has gone unnoticed until now. 

\subsection{Candidate Polarigenic Mechanisms}

Broadly speaking, for a non-magnetic \citep{Grunhut10}, non-binary, early-type star like Deneb, intrinsic broadband linear polarization is produced by electron scattering, either at the photosphere via distortion of the stellar disc; or above it by scattering from an asymmetric gas medium. For the polarization to be variable either the symmetry or the strength of the scattering process must be changing. The two main candidates for a variable polarization are clumpy winds and non-radial stellar pulsations.

\subsubsection{Polarization from Winds}

Polarization variations are most easily explained as arising from scattering in a clumpy stellar wind. Hydrogen gas structures are a ubiquitous source of polarization measuring hundreds to thousands of ppm in early-type systems, such as close binaries and Be stars. Similarly large polarizations are produced in the winds of Wolf-Rayet (WR) stars, where polarimetry is used to study the wind structure \citep{Carmelle89, Moffat91}. Polarization from this mechanism shows no preferred orientation, and though the timescale of polarimetric and photometric variability match, the signals are at best weakly correlated; the ratio of polarimetric to photometric amplitudes, $\sigma_p/\sigma_F$, is $\sim$1/20.

Supergiants show polarimetric variability up to thousands of ppm \citep{Coyne71}. As in WR stars, in earlier type supergiants this variability is attributed to clumpy winds \citep{Hayes84, Lupie87, Bailey24}. Deneb's polarization fluctuations are similar in scale but slower than seen in those stars. We don't yet have enough data for a robust frequency analysis, but there is no clearly favoured polarization direction nor persistent periodicity. 

Similar behaviour is seen in B8\,Ia Rigel; \citet{Hayes86} described its polarization as aperiodic and slowly varying, with no preferred direction nor centroid. Like Deneb, Rigel exhibits semi-periodic light curve fluctuations and episodic H$_\alpha$ features, which \citeauthor{Hayes86} took as supporting evidence for ``temporally and spatially variant mass loss.'' If this is the correct interpretation for both stars then we expect to see irregular photometric variability and features in H$_\alpha$ concordant with large polarization changes.

\begin{figure*}[ht!]
\centering
\includegraphics[width=0.98\textwidth, trim={0 0.3cm 0 0.2cm}, clip]{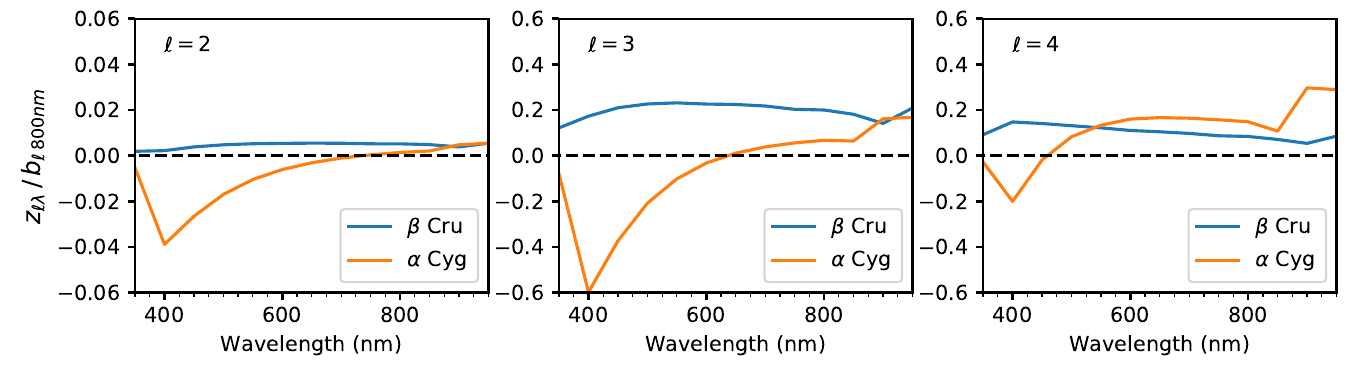}
\includegraphics[width=0.98\textwidth, trim={0 0.3cm 0 0.2cm}, clip]{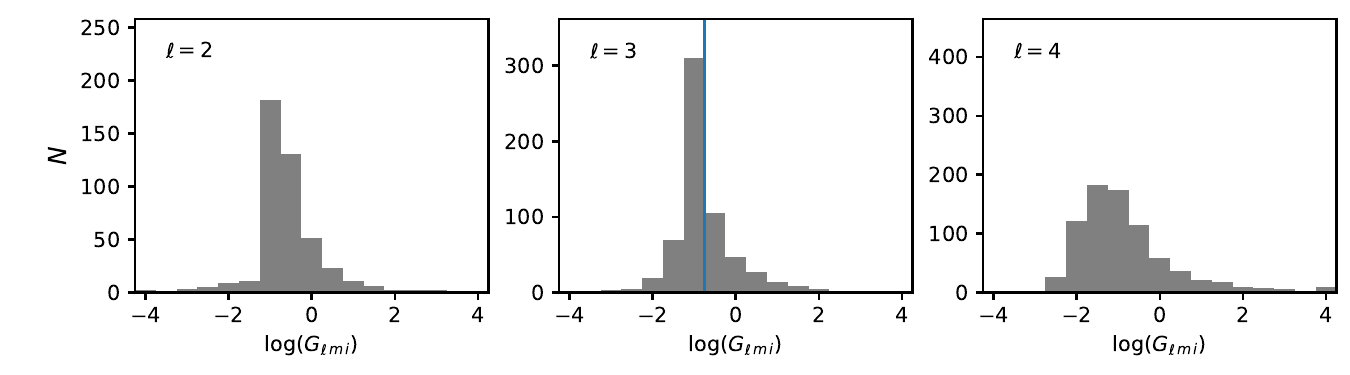}
\caption{Following \citet{Watson83}'s analytical model, the ratio of the polarimetric to photometric pulsation amplitudes is equal to $z_\ell/b_\ell \times G_{\ell m i}$. In the upper panels is the ratio of the photometric scaling factor, $b_\ell$, at 800~nm ($\sim \textit{TESS}~\lambda_{\rm eff}$) and the polarimetric scaling factor, $z_{\ell\lambda}$ for mode degree, $\ell=$ 2 (left), 3 (centre) and 4 (right) for both Deneb ($T_{\rm eff}=$ 8500~K and $\log{g}=$ 1.100 dex) and $\beta$~Cru. In the lower panels are histograms showing the distribution of values of the geometric factor, $G_{\ell m i}$, in which $q$ and $u$ amplitudes are combined as the root sum of the squares. One model instance, $N$, is calculated for each inclination, $i$, between 1$^\circ$ and 89$^\circ$ in one degree increments, and each $m=-\ell,...,0,...,+\ell$ for $\ell=$ 2 (left), 3 (centre) and 4 (right). The vertical blue line indicates $G_{3, -3, 46^\circ}$ -- corresponding to the largest polarimetric amplitude detected for $\beta$~Cru.
\label{fig:model_calcs}}
\end{figure*}

\subsubsection{Non-radial Pulsations}
\label{sec:model}

Neither radial nor dipole pulsations produce polarization, but photospheric distortions caused by non-radial pulsations with mode degree $\ell\ge2$ can produce variable polarization \citep{Odell79}. If this mechanism is at work, then polarimetry, photometry and RV variations will all be correlated with fixed ratios between $q$ and $u$, and the other measurements. In light of the prominent RV work looking at pulsations in Deneb, we present calculations of the polarization to expect from this mechanism here.

\citet{Watson83}'s analytical model enables polarimetric pulsation amplitudes to be determined by multiplying the photometric amplitude by the ratio of scaling factors, $z_{\ell\,\lambda}/b_{\ell\,\lambda}$, and a term derived from the mode geometry we label $G_{\ell m i}$. The terms $b_{\ell\,\lambda}$ and $z_{\ell\,\lambda}$ are calculated from the intensity and polarization dependence on viewing angle, determined from stellar atmosphere models as described by \citet{Cotton22a}, for a given mode degree, $\ell$, and wavelength, $\lambda$. $G_{\ell m i}$ -- which is independent of stellar type -- is a function of $\ell$, azimuthal order, $m$, and inclination, $i$ (see \citealp{Cotton22a} Fig. 4). 

In the top panels of Fig.\,\ref{fig:model_calcs} we have calculated a range of $z_{\ell\,\lambda}/b_{\ell\,\lambda}$ values for both Deneb and $\beta$~Cru. In the lower panels the distribution of $G_{\ell m i}$ is shown for a representative grid of geometries. Together the product of upper and lower panels shows that polarimetric pulsation amplitudes will typically be $\sim$1000 times smaller than photometric, but that in a few rare cases they might actually be larger (i.e. where $N>0$ in the lower panel for values of $G_{\ell m i}$ exceeding the inverse of any point on the line in the corresponding upper panel).

The only polarimetric pulsation detections to date are in $\beta$~Cru \citep{Cotton22a}, which is presented for comparison in Fig. \ref{fig:model_calcs}. Polarization semi-amplitudes of up to $\approx$10 ppm were produced in $\beta$~Cru from a mode with an amplitude 30-40$\times$ smaller than the largest noted in Deneb, so larger effects are possible even though the most likely variability is of a similar scale. The calculations also show that multi-band observations are likely to be a useful diagnostic since the amplitudes will be different, and in some cases -- where $z_{\ell}/b_{\ell}$ changes sign -- of opposite phase.

\subsection{Photometric Frequency Analysis}

We performed a straightforward frequency analysis of the \textit{TESS} data, combining all three sectors, applying a simple linear detrending to the entire time series, and using a $4\times$ oversampled DFT. The result, presented in Fig.\,\ref{fig:gts}(c), reveals periodicities typical of past eras. The prominent peak at 11-12~d corresponds to that first seen by \citet{Paddock35}, others are seen at 18-20~d and $\sim$38~d. Though time series length makes the latter less significant, it is consistent with the $\approx$40~d period identified in H$_{\rm \alpha}$ by \citet{Richardson11}, half the $n\times$72.4~d resumption intervals described by \citet{Abt23}, and double the 18-20~d peak (also about that seen by \citealp{Lucy76, Rzaev23}). If there was a large polarization signal from non-radial pulsations we would expect it to manifest at one of these three detected periods, but none is obvious.

The distinct periods are embedded on a red noise background (i.e. increasing power to longer periods). This is significant since it represents stochastic variability, which could be the counter-part photometric signal to polarization induced by clumpy winds. The photometric to polarimetric amplitude ratio, $\sigma_p/\sigma_F\sim1/11$ is larger but of the same order as that seen in WR stars. 

\pagebreak
\subsection{Alignment with Photometric and Spectroscopic Variability}

In Fig.\,\ref{fig:gts}(a) the time series data is divided into segments, labelled `S[n]' corresponding to the 72.4~d resumption event period; zero marks the last such event identified by \citet{Abt23} at 2459810 BJD. There is no polarimetric data before this, our observations begin 5~d later. About a week after that $q$ begins trending upward reaching its most extreme value of $\approx$3600~ppm at the next photometric minimum, whereupon the timing of local maxima in $q$ are correlated with subsequent photometric minima. The polarimetric amplitude is attenuated much more strongly after the initial outburst than is the photometric signal\footnote{It is also noteworthy that the photometric signal persists beyond the 34\,$\pm$\,1~d duration found in RV by \citet{Rzaev23}.}, such that by $+$50~d no correlation with an 11-12~d cycle is evident. 

The large polarization event cannot have been produced by non-radial pulsations since in that case the photometric to polarimetric amplitude ratio would be constant. However, the two signals seem at least partly related, which suggests that radial (or dipole) pulsations may not be directly responsible for the 11-12~d photometric signal either. The H$_\alpha$ line profile at $+$10~d displays clear additional blue (and possibly red) shifted absorption in the wings of the \mbox{P-Cygni} profile (Fig.\,\ref{fig:gts}(b)); a likely explanation is the ejection of a large gas clump. In subsequent spectra, similar absorption features (in H$_\alpha$ and/or H$_\beta$) are only present at $+$361 and $+$362~d (at the S5/S6 boundary) where the polarization also deviates strongly from the median. In this case the features are stronger -- both blue and red shifted -- perhaps indicative of closer proximity to a smaller event. If both events do correspond to ejections then they propagate in different directions.

\begin{table*}[]
\begin{minipage}[l]{0.74\textwidth}
\centering
\tabcolsep 3.5 pt
\begin{tabular}{lccrrrrrrrr}
\hline
Fil.                  & \multicolumn{1}{c}{$\lambda_{\rm eff}$} &  \multicolumn{1}{c}{Date range}          & \multicolumn{1}{c}{$n$}     & \multicolumn{1}{c}{$\Delta q$} & \multicolumn{1}{c}{$\Delta u$} & \multicolumn{1}{c}{$\sigma_{\Delta q}$} & \multicolumn{1}{c}{$\sigma_{\Delta u}$} & \multicolumn{1}{c}{$\bar{\rm Err.}$}   &   $\bar{p}_{\, \rm imp}$ & $\bar{\theta}_{\rm imp}$\\    
                        & \multicolumn{1}{c}{(nm)} &  \multicolumn{1}{c}{JD$-$2459810}  &       & \multicolumn{1}{c}{(ppm)} & \multicolumn{1}{c}{(ppm)}      & \multicolumn{1}{c}{(ppm)} &   \multicolumn{1}{c}{(ppm)}   & \multicolumn{1}{c}{(ppm)} & \multicolumn{1}{c}{(ppm)} & \multicolumn{1}{c}{($^\circ$)}\\
\hline
$u^\prime$   &   377    &   33 --  \phantom{0}98  & 11    & $+$300.1    &  $-$164.3   & 53.5    & 113.1 & 55.9  & 3852.7 & 31.75 \\
$g^\prime$   &   468    & \phantom{0}5 -- 416 & 88 &&&&&& 3881.0 & 34.28  \\
$r^\prime$   &   614    &   33 --  \phantom{0}66  & 10    & $-$221.2    &  $-$143.7   & 66.2    &  90.7 & 27.1  & 3669.5 & 35.48 \\
$i^\prime$   &   762    &   66 --  \phantom{0}98  &  2    & $-$250.3    &  $-$515.9   & 52.2    &  19.1 & 31.0  & 3309.6 & 34.66 \\
$z^\prime$   &   892    &   68 -- 117             & 18    & $-$551.9    & $-$1119.9   & 98.6    &  63.5 & 48.4  & 2639.0 & 35.41 \\
\hline
\end{tabular}
\end{minipage}
\begin{minipage}[l]{0.24\textwidth}
\caption{Multi-band polarimetry summary. Columns 3-9 correspond to observations concurrent with $g^\prime$ to which the difference is taken; 10-11 are calculated by first adding the median $g^\prime$ $q$ and $u$ values to $\Delta q$ and $\Delta u$ for each other band.  \label{tab:multi}}
\end{minipage}
\end{table*}

No polarization changes as extreme as that in S1 are seen subsequently\footnote{The peak of the extreme event is, however, consistent with the observations of \citet{Hall50} in both $p$ and $\theta$.}, and we have no more photometric data with which to compare. There is insufficient polarimetric data for a meaningful frequency analysis. However, smaller scale changes are noticeable on timescales of $\sim$10-40~d; some are coincident with the segment transitions: Near the S1/S2 transition $p$ is a minimum; at the S4/S5 transition $q$ is a minimum; at the beginning of S6 $u$ begins to decrease. These events are perhaps easiest to identify in $\theta$ and seem to correspond to inflection points rather than abrupt changes. The pattern is somewhat subjective, but if it persists it might be indicative of a deeper process that propagates through to the photosphere and drives the winds. Such a hypothesis would be best tested with simultaneous spectroscopic (incl. RV) and polarimetric data with nightly cadence. The line cores of H$_\alpha$ and H$_\beta$ shift relative to lines more representative of the photosphere in Fig.\,\ref{fig:gts}(b), however our spectroscopic data is not extensive enough, as yet, to search for meaningful periodicities or connections. 

\subsection{Wavelength Dependence}
\label{sec:multi}

On 28 occasions observations were made in one or more other SDSS bands alongside $g^\prime$ with PICSARR (within 35 mins). In Tbl.\,\ref{tab:multi} we summarize these in terms of the difference $\Delta[q/u] = [q/u]_{fil} - [q/u]_{g^\prime}$ in order to examine wavelength dependence independent of the dominant trends in polarization. To first order the multi-band observations are offset from but otherwise follow the same trends as those evident in $g^\prime$ -- i.e. the standard deviations in $\Delta q$ and $\Delta u$ are mostly 1-2$\sigma$. Only for $r^\prime-g^\prime$ is $\sigma_{\Delta u} > 3\sigma$. However, the nominal errors for the $u^\prime$ and $z^\prime$ bands are larger, so those measurements are not as sensitive (partly due to the camera quantum efficiency curve, but also because of shorter non-$g^\prime$ exposures, so future improvement will not be \mbox{difficult}).

Whether the variable component of the polarization is wavelength independent or not is important, since this will discriminate between optically thin wind structures (independent) and either optically thick ones or another mechanism (dependent). Non-radial pulsations could be such a mechanism. Dual filter observations such as these will account for an optically thin wind if it is insubstantial enough to see through to the photosphere. The scale of variability reflected in Tbl.\,\ref{tab:multi} is consistent with what we could expect in this case from the pulsation calculations in Sec.\,\ref{sec:model}; any differential pulsation signal is constrained to be $\lessapprox$100~ppm. So far the multi-band data does not reflect the expected difference in sign (i.e. phase), but it is too sparse to draw conclusions yet.

In Tbl.\,\ref{tab:multi} $p$ and $\theta$ are the values implied by adding the median $g^\prime$ $q$ and $u$ values to the mean band differences. A 3-parameter \citeauthor{Serkowski68} Law (\citeyear{Serkowski68}) fit to the individual data points modified in the same way (incl. 29 common $g^\prime$ points) gives the maximum interstellar polarization $p_{\rm max}=3898\,\pm\,9$~ppm, at a wavelength $\lambda_{\rm max}=464\,\pm\,6$~nm, with ``constant'' $K=0.87\,\pm\,0.04$. The exact value of $\lambda_{\rm max}$ is sensitive to the assumed (median) $g^\prime$ value, but regardless is bluer than the typical Galactic value of 550~nm \citep{Serkowski75}, which implies smaller grains and/or another component. 

The Serkowski fit reduced $\chi^2$ is only 4.35: $p$ is high in $u^\prime$ and low in $z^\prime$. Assuming the adopted (median) $g^\prime$ values are truly representative -- and if, as \citet{Hayes86} reports for Rigel, there is no clear centroid, it might not be -- together with the complex $\theta(\lambda)$ behaviour, this indicates multiple constant polarization components. \citet{Behr59b}'s explanation -- two or more distinct dust clouds on Deneb's sight line -- seems likely given the star's proximity on the sky to the Pelican nebula, but other contributions, such as from an asymmetric wind (as suggested by interferometry, \citealp{Chesneau10}), are not ruled out. \citet{Coyne71} thought similar behaviour in $p(\lambda)$ observed for other supergiants was intrinsic in origin.

\section{conclusions} 
\label{sec:conclusions}

Extraordinarily, the broadband polarization observations of Deneb reported here are the first such measurements in about 60 years. They reveal conclusively for the first time that the star is a large amplitude polarization variable. The median polarization in $g^\prime$ is $p=$\,3881.0~ppm and $\sigma_p=$\,687.0~ppm, about 1/11th of the flux variability.

An event producing large polarization occurred shortly after the last pulsation resumption event identified by \citet{Abt23} in \textit{TESS} photometry. Deneb's dominant polarization signature appears related to its photometric variations in a way that probably indicates structures in its winds or extended atmosphere, which would make asteroseismology difficult. However, the presence of a smaller signal due to pulsation is not ruled out; future multi-band polarimetric observations will provide an invaluable diagnostic in this respect. We plan such observations concurrent with spectroscopy to correspond with upcoming \textit{TESS} photometry. \citet{Abt23} called for more photometry and spectroscopy of Deneb; polarimetry is also critical to understanding this enigmatic object.

\begin{acknowledgments}
We thank the Friends of MIRA for their support, \mbox{Normandy} Filcek for observing assistance, Sarbani Basu for help acquiring a reference, and \mbox{Wm. Bruce} Weaver for bringing our attention to \citet{Abt23} as well as useful comments on the manuscript.

\textit{Data Statement:} The raw \textit{TESS} data used in this paper can be found at \dataset[DOI: 10.17909/w1wk-mn94]{https://doi.org/10.17909/w1wk-mn94}
\end{acknowledgments}

\bibliography{ms}{}
\bibliographystyle{aasjournal}

\end{document}